\documentclass[prx,aps,twocolumn,amsmath,amssymb,reprint,numbers,superscriptaddress,noeprint,longbibliography]{revtex4-2}

\usepackage{tikz}
\usepackage{hyperref}

\begin{document}

\title{Identification of the I$_{10}$ Donor in ZnO as a Sn–Li Complex \\ with Large Hyperfine Interaction}

\author{Xingyi Wang}
\affiliation{Department of Electrical and Computer Engineering, University of Washington, Seattle, Washington 98195, USA}
\author{Sai Mu}
\affiliation{Department of Physics and Astronomy and SmartState Center for Experimental Nanoscale Physics, University of South Carolina, Columbia, SC, 29208, USA}
\affiliation{Materials Department, University of California, Santa Barbara, California 93106-5050, USA}
\author{Jeong Rae Kim}
\affiliation{Department of Applied Physics and Materials Science, California Institute of Technology, Pasadena, California 91125, USA}
\author{Ethan R. Hansen}
\affiliation{Department of Physics, University of Washington, Seattle, Washington 98195, USA}
\author{Yaser Silani}
\affiliation{Department of Physics, University of Washington, Seattle, Washington 98195, USA}
\author{Lasse Vines}
\affiliation{Department of Physics/Centre for Materials Science and Nanotechnology, University of Oslo, Blindern, Oslo N-0316, Norway}
\author{Joseph Falson}
\affiliation{Department of Applied Physics and Materials Science, California Institute of Technology, Pasadena, California 91125, USA}
\author{Chris G. Van de Walle}
\affiliation{Materials Department, University of California, Santa Barbara, California 93106-5050, USA}
\author{Kai-Mei C. Fu}
\affiliation{Department of Electrical and Computer Engineering, University of Washington, Seattle, Washington 98195, USA}
\affiliation{Department of Physics, University of Washington, Seattle, Washington 98195, USA}
\affiliation{Physical Sciences Division, Pacific Northwest National Laboratory, Richland, Washington 99352, USA}

\email{kaimeifu@uw.edu}

\date{\today}

\begin{abstract}

Donor impurities in wide direct band gap semiconductors provide a promising platform for spin–photon quantum technologies by combining a donor spin qubit with optically addressable transitions. In ZnO, the shallow donor with the largest reported binding energy has long been associated with the I$_{10}$ bound exciton line, but its microscopic origin has remained unresolved. Here we demonstrate the controlled formation and identification of this donor as a Sn–Li complex through a combination of ion implantation, annealing, optical spectroscopy, and first-principles calculations. Resonant two-laser coherent population trapping measurements reveal an electron–$^{119}$Sn hyperfine interaction of $392 \pm 15$\,MHz, establishing a coupled electron-spin-1/2, nuclear-spin-1/2 system with one of the largest hyperfine couplings reported for shallow donors in semiconductors. Density functional theory calculations show that a nearest-neighbor Sn$_{\mathrm{Zn}}$–Li$_{\mathrm{Zn}}$ complex has favorable formation energetics, donor character with the electron localized on Sn, and an extrapolated hyperfine interaction consistent with experiment. The large donor binding energy and excited-state structure indicate enhanced thermal robustness of the optical transition relative to conventional group-III donors, while the strong hyperfine interaction enables fast electron–nuclear spin control and prospects for direct nuclear-spin–photon interfaces. We further observe efficient optically induced nuclear spin polarization, highlighting a path toward nuclear spin initialization. More broadly, our results reveal how a donor–acceptor complex can access previously unexplored regimes of shallow donor physics, extending the design space of quantum defects beyond isolated substitutional dopants.

\end{abstract}

\maketitle

\section{\label{sec:level1}Introduction}

Donor impurities in semiconductors provide a versatile platform for quantum information science, combining long-lived spin degrees of freedom with well-established materials control. In particular, donor-bound electron spins can serve as qubits that are naturally coupled to nearby nuclear spins through the hyperfine interaction, enabling multi-qubit registers with exceptional coherence~\cite{Edlbauer_2025}. In silicon, donor systems have achieved remarkable performance~\cite{McCallum_2021}; however, the indirect band gap of Si limits the efficiency of optical spin–photon interfaces, constraining applications in quantum networking and distributed quantum information processing.

Wide direct-band-gap semiconductors offer an attractive alternative host by enabling radiatively efficient optical transitions while retaining the favorable spin properties of shallow donors. Among these materials, ZnO has recently emerged as a promising host for donor-based spin–photon quantum technologies. The donor-bound exciton in ZnO provides a direct optical interface to the electron spin, and long electron spin relaxation times exceeding 0.5 s have been demonstrated~\cite{Niaouris_2022}. Progress toward deterministic donor creation through ion implantation and annealing~\cite{Wang_2023}, as well as isolation of individual donors~\cite{Hansen_2024}, further establishes ZnO as a viable quantum defect platform. Nevertheless, the available donor species in ZnO remain limited, and identifying donors with improved optical and spin properties is an important outstanding challenge. 

\begin{figure}[b]
    \centering
    \includegraphics[width=1.\linewidth]{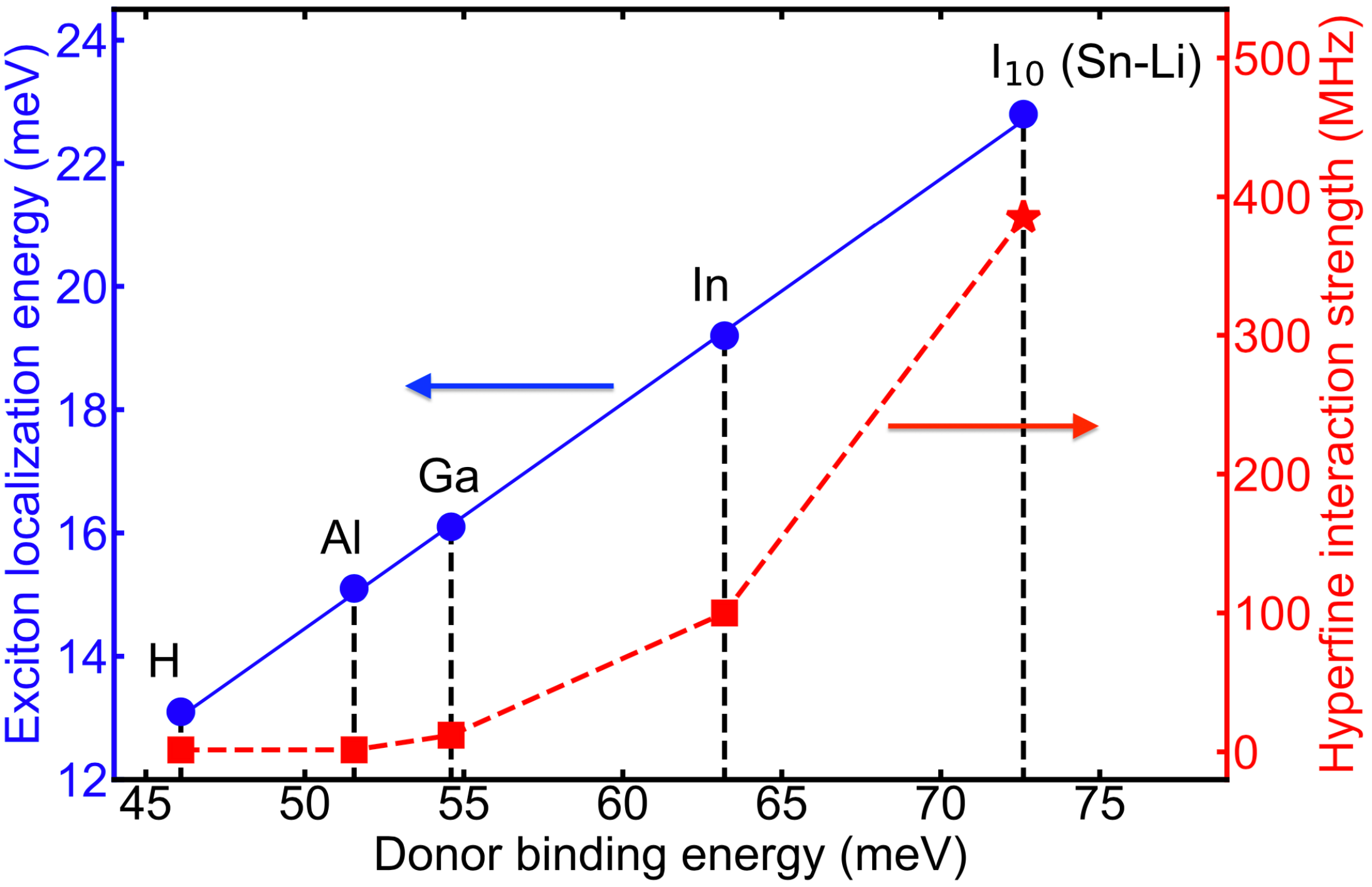}
    \caption{Dependence of bound exciton localization energy (blue) and hyperfine interaction strength (red) as a function of donor binding energy. Circles and squares are Ref.~\onlinecite{Meyer_2004}. Star represents this work.}
    \label{fig:binding}
\end{figure} 

A key parameter governing both optical robustness and spin functionality is the donor binding energy. As illustrated in Fig.~\ref{fig:binding}, larger binding energies increase the localization energy of the donor-bound exciton, pushing optical transitions further from the band edge and enabling operation at elevated temperatures. At the same time, the hyperfine interaction between the donor electron and nucleus generally increases with donor atomic number, providing faster electron–nuclear spin control and potentially enabling direct nuclear-spin–photon entanglement when the hyperfine splitting exceeds the optical linewidth. These considerations motivate the search for donor systems in ZnO with binding energies exceeding those of conventional group-III substitutional donors. 

The shallow donor with the largest reported binding energy (73\,meV) is associated with the I$_{10}$ donor bound exciton line, whose microscopic origin has remained unresolved despite decades of study. It is associated with a substitutional Sn at a Zn site, a double donor in ZnO~\cite{Cullen_2013}. Previous work suggests involvement of a compensating acceptor such as Li or Na ~\cite{Meyer_2004,Abbas_2021}. However, direct experimental identification of the defect complex responsible for the I$_{10}$ transition and characterization of its spin properties have been lacking.

In this work we demonstrate the controlled formation of the I$_{10}$ line via implantation and annealing of Sn and Li in ZnO, supporting the identification of the I$_{10}$ as a Sn double donor-Li acceptor complex. Two-laser spectroscopy reveals two coherent population trapped states corresponding to the spin-1/2 Sn nucleus with an extremely large hyperfine interaction $A_{\mathrm{exp}} = 392 \pm 15$\,MHz.  First-principles calculations confirm that $\mathrm{Sn_{Zn}-Li_{Zn}}$ complex, in which Sn and Li occupy adjacent Zn sites, exhibits favorable formation energetics, donor character with the electron localized on Sn, and a hyperfine interaction consistent with experiment. We further observe efficient optically induced nuclear spin polarization, highlighting opportunities for nuclear spin initialization.

\section{\texorpdfstring{Formation of I$_{10}$ donor in ZnO}{Formation of I10 donor in ZnO}}

\begin{figure*}
    \includegraphics[width=1\linewidth]{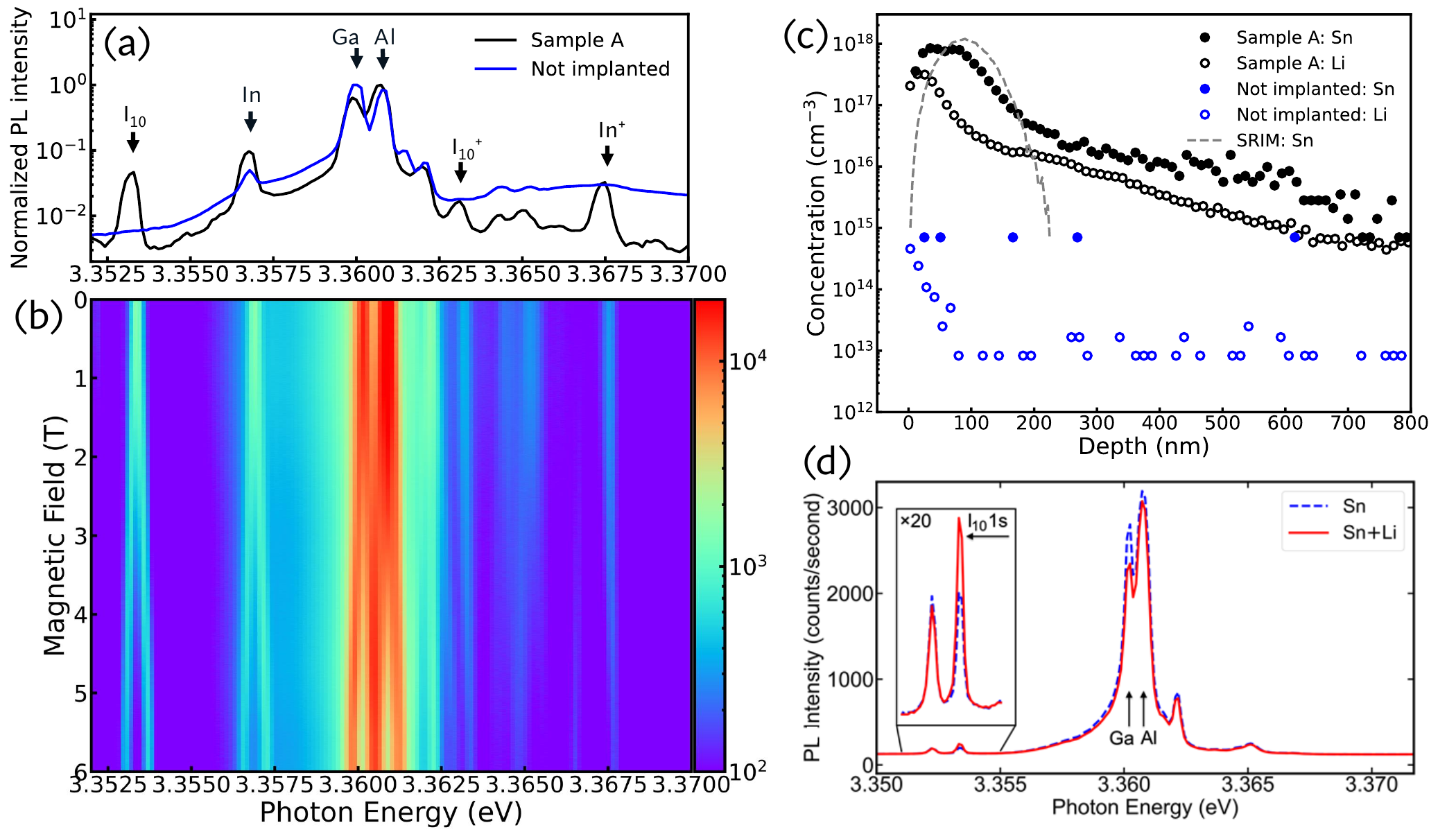}
    \caption{(a) Zero-field photoluminescence spectrum of Sn-implanted sample A and a non-implanted piece of the same substrate. $T= 7$\,K. (b) Magneto-PL spectrum of sample A. T=8.9\,K. The color bar shows PL intensity in counts/seconds. (c) SIMS measurement of the Sn and Li density in sample A and an unimplanted piece of the same substrate. In the unimplanted sample, Sn is at the detection limit and Li deeper than 100\,nm is at the detection limit in the non-implanted sample. (d) Formation of I$_{10}$ (Sn-Li) with and without Li implantation. Li implantation and annealing showing enhanced formation with Li implantation. $T=9.5$\,K. }
    \label{fig:form}
\end{figure*}

We first demonstrate the formation of the I$_{10}$ line via Sn implantation followed by high-temperature annealing. In Sample A, $^{119}$Sn (100\% spin-1/2 isotope) is implanted at energy 380\,keV and fluence of 10$^{13}$\,ions/cm$^2$ into a 1.7\,\textmu m-thick MBE epilayer grown on a hydrothermal ZnO (0001) substrate (Tokyo Denpa) (Appendix~\ref{app:growth}). Following implantation, the sample is annealed at 800\,$^\circ$C in an oxygen atmosphere for one hour. Figure~\ref{fig:form}(a) shows the zero-field photoluminescence (PL) spectrum for sample A and non-implanted sample from the same substrate. Al, Ga and In bound exciton lines are present prior to implantation. Post implantation we observe a new line, the I$_{10}$ line. We also observe an increase in the indium which is attributed to indium diffusion from the hydrothermal substrate to the epilayer. The magneto-PL spectrum (Fig.~\ref{fig:form}b) further confirms the identification of the I$_{10}$ line as a shallow donor bound exciton transition~\cite{Rodina_2004}; I$_{10}$ splits into a doublet with magnetic field (B $\perp$ [0001]), with the splitting dominated by the donor electron Zeeman splitting. Similar to the ionized In$^+$ line, the ionized I$_{10}^+$ does not split with field~\cite{Rodina_2004}. 

Secondary ion mass spectroscopy (SIMS) measurements provide insight into how I$_{10}$ forms with only Sn implantation. As shown in Fig.~\ref{fig:form}(c), the Sn distribution follows the predicted simulated ion implantation profile (SRIM~\cite{Ziegler_2010}), with some modest diffusion observed. Prior to annealing, Li and Sn are close to the measurement detection floor. Post anneal, we find that the Li profile appreciably overlaps with the implanted Sn profile, supporting that Li has diffused from the substrate into the epilayer. 

In a second sample, we demonstrate the enhanced formation of I$_{10}$ with combined Sn and Li implantation. In Sample B $^{119}$Sn (380\,keV and fluence of 10$^{12}$\,ions/cm$^2$) is implanted into a 1.5\,\textmu m-thick MBE epilayer, again grown on a hydrothermal ZnO (0001) Tokyo Denpa substrate). After annealing (780\,$^\circ$C, 2 hours in oxygen), the sample is implanted with $^7$Li through a grid mask (120\,keV,  10$^{11}$ ions/cm$^2$) and annealed at 600\,$^\circ$C in oxygen for 20 minutes. Fig.~\ref{fig:form}(d) shows a comparison between a region implanted only with Sn versus a region implanted with both Sn and Li. A clear enhancement of the I$_{10}$ is observed in the Li-implanted regions. In contrast, a slight decrease in the other band edge exciton lines is observed which is attributed to some residual lattice damage due to the Li implantation. This measurement further confirms the involvement of Li in the complex.

\section{\texorpdfstring{Observation of electron-$^{119}$Sn hyperfine interaction}{Observation of electron-119Sn hyperfine interaction}}

\begin{figure*}
    \centering
    \includegraphics[width=1\linewidth]{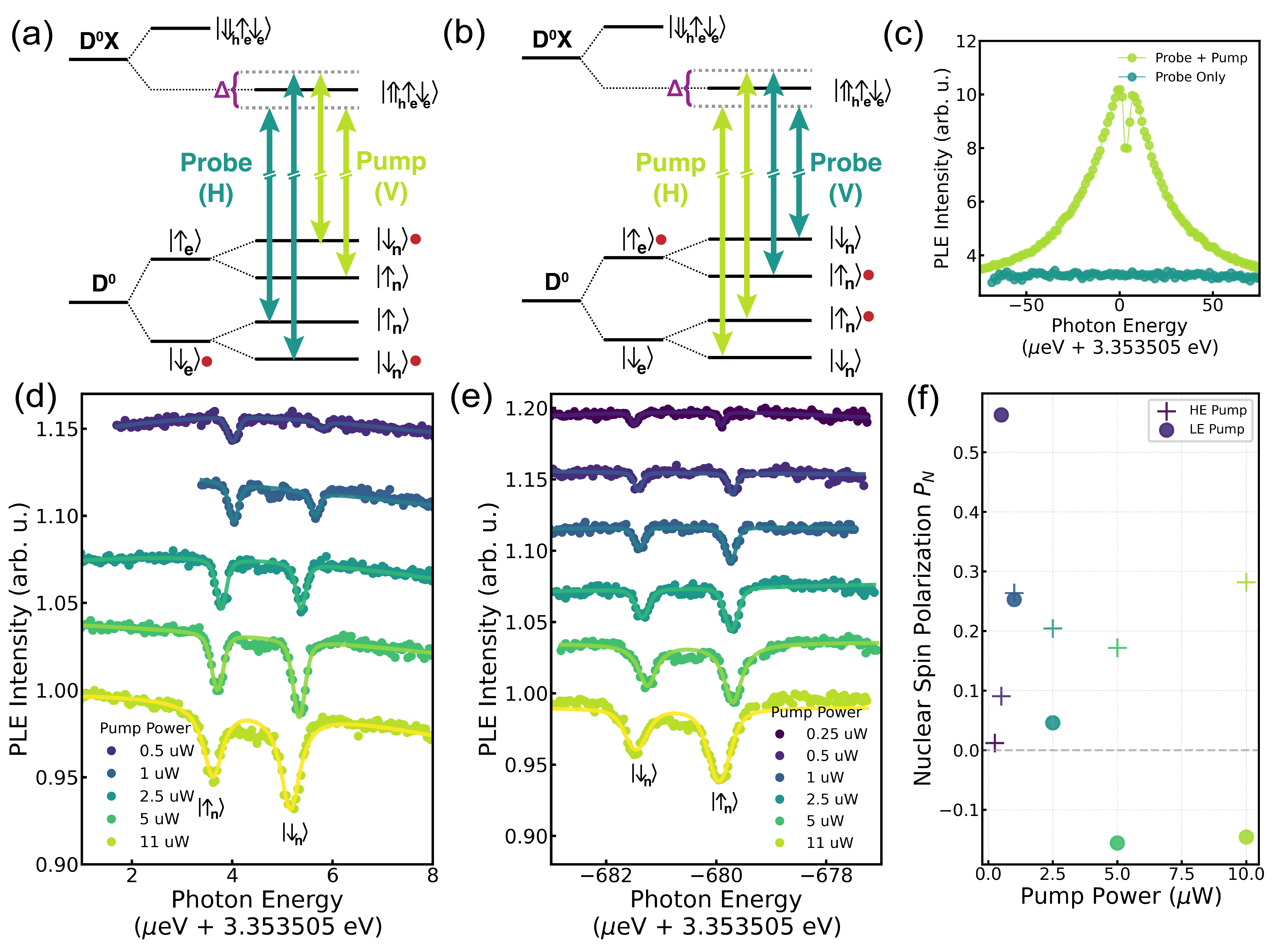}
    \caption{Experimental energy configurations in the (a) low (LE) energy fixed-frequency pump case and (b) high-energy (HE) fixed-frequency pump case. (c) Photoluminescence excitation probe scan over the high-energy transition with and without the pump excitation. (d) and (e) Power series of the pump laser for (a) LE pump and (b) HE pump, respectively. (f) Nuclear spin polarization corresponding to (d) and (e). All scans are taken at B=6\,T and T=8\,K.}
    \label{fig:CPT}
\end{figure*}

We next perform high-spectral resolution photoluminescence excitation spectroscopy on the I$_{10}$ line to elucidate potential fine structure in the resonance. We focus on Sample A due to its relative brightness. Under optical resonant excitation of I$_{10}$ at zero magnetic field, additional excitonic transitions can be observed; an excited donor bound exciton state I$_{10}^*$ and satellite transitions to excited donor states. The I$_{10}^*$  is observed at 4.2\,meV above the main transition~\ref{app:I10*}. This large energy splitting indicates the Sn-Li optical linewidth will be more robust to temperature than  Al, Ga, and In. The satellite transitions correspond to relaxation from the donor bound exciton to an excited donor state, as well as phonon replicas of the main transition (Appendix~\ref{app:replicas}). In photoluminescence excitation (PLE) spectroscopy, we utilized these satellite transitions as a probe of the excited state population under resonant excitation~\cite{Wang_2023}. Curiously, resonant excitation of the I$_{10}$ line leads to extremely sharp features near the Ga and Al replicas (Fig.~\ref{fig:TES}) . While the mechanism for this excitation in unclear, these transitions also serve as probe for I$_{10}$ absorption.

The ensemble linewidth of the I$_{10}$ line ranges between 6-16\,GHz, prohibiting the observation of any hyperfine features. We thus perform two-laser coherent population trapping to resolve the hyperfine interaction between the donor electron and $^{119}$Sn nucleus. The pump and scanning probe laser are illustrated in Fig.~\ref{fig:CPT}(a). The pump laser is at a fixed energy on resonance with the electron spin $|\uparrow\rangle$ transition that couples both nuclear spin states with one excited state. The excited state is not expected to exhibit a significant hyperfine splitting due to the spin-singlet nature of the two electrons and the p-like nature of the bound hole so the nuclear spin states are treated as degenerate. The probe laser scans across the transition between the electron spin $|\downarrow\rangle$ state and the same excited state. Two-photon resonance and thus coherent population trapping occurs at the two frequencies which conserve nuclear spin. The difference between these frequencies is the hyperfine constant. The opposite pump-probe configuration is depicted in Fig.~\ref{fig:CPT}b. Due to the negative gyromagnetic ratio of $^{119}$Sn, the hyperfine interaction is negative, resulting in the less common energy ordering of the ground state manifold.

A coarse scan with only the probe excitation beam is shown in Fig.~\ref{fig:CPT}(c). At 6\,T field, no absorption transition is observed due to efficient optical pumping into the opposite electron spin state. We next apply the fixed pump. The transition is now visible and a single CPT dip is observed. High resolution scans across this dip reveal the hyperfine structure. A series of scans are taken in which the pump power is varied with the pump on either the low energy transition (Fig.~\ref{fig:CPT}d) or the high energy transition (Fig.~\ref{fig:CPT}e). The first observation is that in all scans, two dips are observed separated by $392 \pm 15$~MHz. The doublet structure indicates that the hyperfine interaction is dominated by the spin-1/2 $^{119}$Sn nuclear spin and not the Li; the dominant isotope of Li is spin-3/2. Curiously, though, the amplitude of the dips are not equal, indicating that the nuclear spin is polarized. Moreover, the amplitude is dependent on the relative optical intensity of the pump and probe.

From Fig.~\ref{fig:CPT}(a), we see that the low energy probe dip corresponds to the $|\uparrow_n\rangle$ nuclear spin state while the high energy dip corresponds to the $|\downarrow_n\rangle$ state. Thus as the pump power increases (moving vertically down in Fig.~\ref{fig:CPT}d), which increases the electron $|\downarrow_e\rangle$, we observe an increase in the $|\downarrow_n\rangle$ population.  These states are marked with a red dot. The opposite is true in the configuration corresponding to Fig.~\ref{fig:CPT}(b); a stronger pump will correspond to population increase for $|\uparrow_n\rangle$, which also corresponds to an increase in the high energy dip. While it is difficult to make a quantitative comparison between the two cases due to the different pump and probe beam shapes and potentially different overlaps between the experiments, in both cases we observe that the nuclear spin polarization is following the electron spin polarization. The nuclear spin polarization $P_N = (P_{\uparrow_n}-P_{\downarrow_n})/(P_{\uparrow_n}+P_{\downarrow_n})$ can be quite substantial (Fig.~\ref{fig:CPT}f). In the lowest pump case, in which the probe optical pumping is dominating, the polarization exceeds 0.5 which is a 10$^3$-fold enhancement over equilibrium conditions.

In these experiments, the observed nuclear spin polarization is likely governed by two competing mechanisms. First, optical excitation generates a non-equilibrium electron spin polarization, which in turn drives nuclear spin polarization through the Overhauser effect~\cite{Jeffries_1960}. Away from two-photon resonance, this mechanism is expected to dominate. In contrast, at two-photon resonance an additional process becomes relevant: the nuclear spin can be preferentially driven into the state associated with the coherent population trapping (CPT) condition~\cite{Adambukulam_2024, Ethier-Majcher_2017, Togan_2011}. In this regime, electron–nuclear flip-flop processes enable nuclear spin diffusion until population accumulates in the CPT dark state. The competition between these mechanisms provides a natural explanation for the observed trends: higher nuclear polarization at weak pump power, where the Overhauser process dominates, and reduced polarization at strong pump power, where diffusion into the CPT state becomes more efficient. Time-resolved measurements, beyond the scope of the present work, would help quantify the relative efficiencies of nuclear spin pumping under resonant and off-resonant two-laser conditions. 

\section{\texorpdfstring{First-Principles calculation of electron-$^{119}$Sn hyperfine interaction}{\textit{\emph Ab initio} calculation of electron-119Sn hyperfine interaction}}

Density functional theory (DFT) calculations are performed to determine both the formation energy and hyperfine interaction strength for the Sn-Li complex. We adopt the approach developed in Swift \emph{et al.}~\cite{swift2020first,karbasizadeh2024transition} to construct supercells for defect calculations by expanding the unit cell of the host material and extrapolate the shallow donor properties in the dilute limit. The stability of the defects is assessed by calculating their formation energies $(E^f)$ using HSE in a 128-atom supercell (as described in Appendix~\ref{app:dft}). 

\begin{figure}
    \centering
    \includegraphics[width=1.\linewidth]{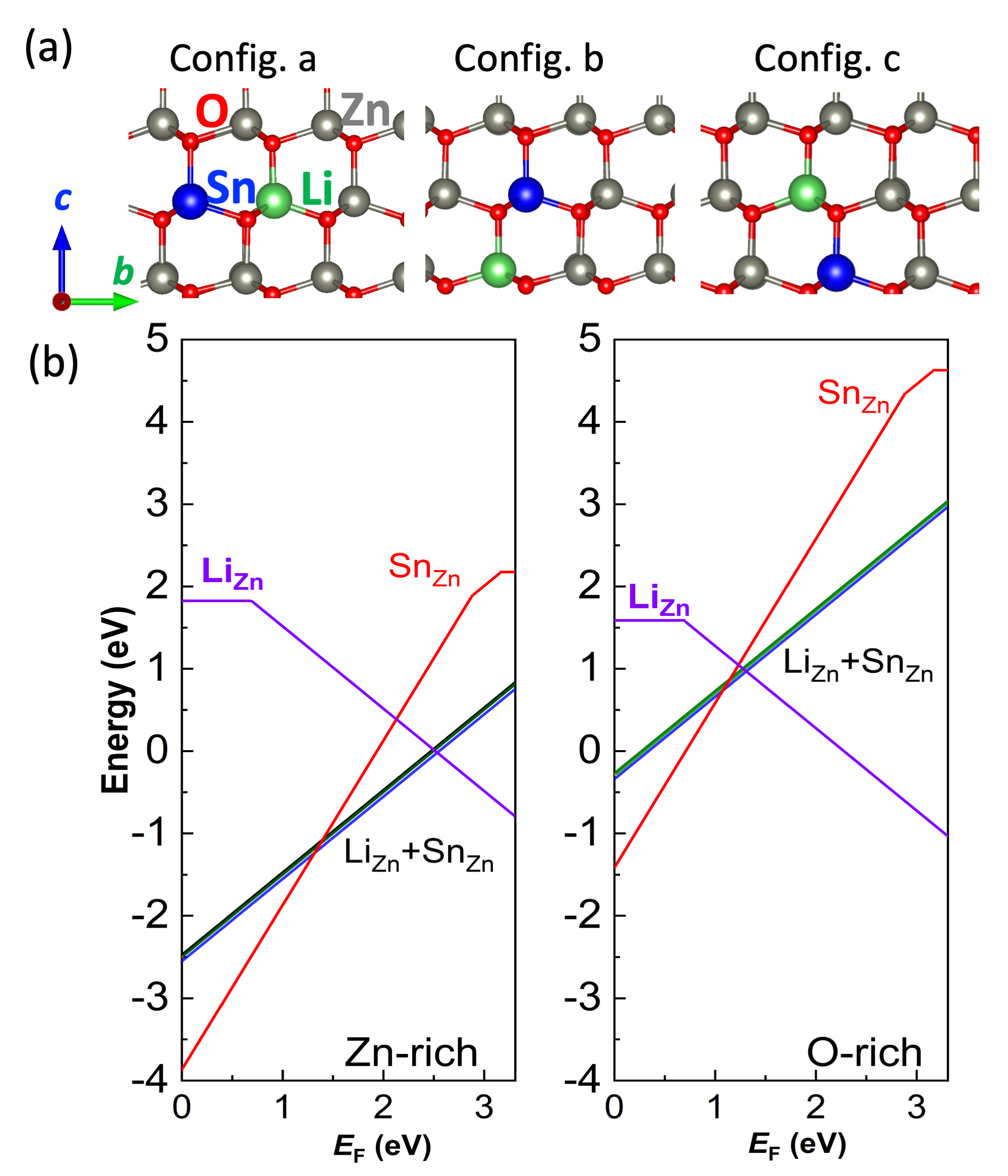}
    \caption{(a) Structural illustration of three symmetry-inequivalent Sn$_\textnormal{Zn}$+Li$_\textnormal{Zn}$ defect complexes (denoted as a, b and c). (b) Formation energy $E^f$ of Li$_\textnormal{Zn}$, Sn$_\textnormal{Zn}$ and Sn$_\textnormal{Zn}$+Li$_\textnormal{Zn}$ defect complexes in ZnO as a function of Fermi level $E_{\mathrm{F}}$ under Zn-rich (left panel) and O-rich conditions (right panel). Three Sn$_\textnormal{Zn}$ + Li$_\textnormal{Zn}$ defect complexes are considered, and they exhibit very similar, nearly overlapping formation energies.}
    \label{fig:ef}
\end{figure}

We first consider substitutional Li on the cation site (Li$_{\mathrm{Zn}}$) in ZnO. The calculated formation energies under Zn-rich and O-rich conditions are shown in Fig.~\ref{fig:ef}, in good agreement with previous calculations~\cite{Carvalho_2009}. The charge-state transition level (0/$-$) is located 0.69~eV above the valence-band maximum (VBM). For Fermi levels ($E_\text{F}$) lying below this level, Li$_{\mathrm{Zn}}$ is in the neutral charge state, whereas for $E_\text{F}$ above it, Li$_{\mathrm{Zn}}$ acts as a single acceptor. We note that in the experiment, the ZnO samples are $n$-type with a Fermi level near the conduction band.
Under O-rich conditions, the formation energy of Li$_{\mathrm{Zn}}$ is lower, indicating that its formation is more favorable.

For Sn on Zn site (Sn$_{\mathrm{Zn}}$), two charge-state transition levels, (2$+$/$+$) and ($+$/0), are identified at 0.43~eV and 0.14~eV below the conduction-band minimum (CBM), respectively. 
When the Fermi level lies below the (2$+$/$+$) level, Sn$_\textnormal{Zn}$ donates two electrons. For $E_\textnormal{F}$ between the (2$+$/$+$) and ($+$/0) levels, Sn$_\textnormal{Zn}$ donates a single electron.
 It becomes neutral when the Fermi level is above the ($+$/0) level. Unlike Li$_{\mathrm{Zn}}$, the formation of Sn$_{\mathrm{Zn}}$ is more favorable under Zn-rich conditions. 

We next consider the Sn$_{\mathrm{Zn}}$–Li$_{\mathrm{Zn}}$ defect complex (hereafter referred to as Sn–Li for simplicity). Three symmetry-inequivalent configurations of the Sn–Li complex are identified, denoted as a, b, and c [Fig.~\ref{fig:ef} (a)]. Their calculated formation energies are shown in Fig.~\ref{fig:ef} (b). All these Sn–Li complexes behave as shallow donors. Although configuration b is slightly energetically favorable, all three configurations have very similar, nearly overlapping formation energies. Experimentally, only a single donor-bound exciton transition is observed, supporting that the difference in the charge-transition levels (and thus ionization energy) of the three configurations is very small. For $n$-type ZnO samples with $E_\textnormal{F}$ near the CBM, the formation energy of Sn–Li complex is 0.8 eV under Zn-rich conditions and 3.0 eV under O-rich conditions.

To assess the stability of the Sn--Li complex, we calculate its binding energy, defined as
\begin{equation}
\label{eq:bind}
\begin{split}
E_\mathrm{bind}[(\mathrm{Sn}_\mathrm{Zn}-\mathrm{Li}_\mathrm{Zn})^+] 
&= E^f(\mathrm{Sn}_\mathrm{Zn}^{2+}) + E^f(\mathrm{Li}_\mathrm{Zn}^-) \\
&\quad - E^f[(\mathrm{Sn}_\mathrm{Zn}-\mathrm{Li}_\mathrm{Zn})^{+}] .
\end{split}
\end{equation}
The calculated binding energy of the Sn--Li complex is $+1.12$~eV, indicating a stable complex. 

The isotropic hyperfine parameter in SI units is defined as
\begin{equation}
A=\frac{2\mu_0}{3} g_e \mu_B g_I \mu_N 
\int \delta_T(r-R)\,\sigma(r)\,dr,
\end{equation}
where $g_e$ denotes the electron $g$-factor, $\mu_B$ the Bohr magneton, 
$g_I$ the nuclear $g$-factor of the dopant, and $\mu_N$ the nuclear magneton. 
Here, $R$ represents the position of the nucleus and $\sigma(r)$ the spin density. The calculations follow the procedures described in 
Refs.~\onlinecite{Szasz_2013, Blochl_2000, Vandewalle_1993, Yazyev_2005}. 
We note that the spin density of Sn--Li donor state is more centered around the Sn atom rather than on Li, so the hyperfine interaction is dominated by the Sn site.
The isotope $^{119}$Sn is chosen for the calculation of the hyperfine 
parameter of the Sn--Li donor, as it possesses a nonzero 
nuclear spin and the spin density of the Sn-Li donor state is found to be localized on the Sn atom.

\begin{figure}
    \centering
    \includegraphics[width=1\linewidth]{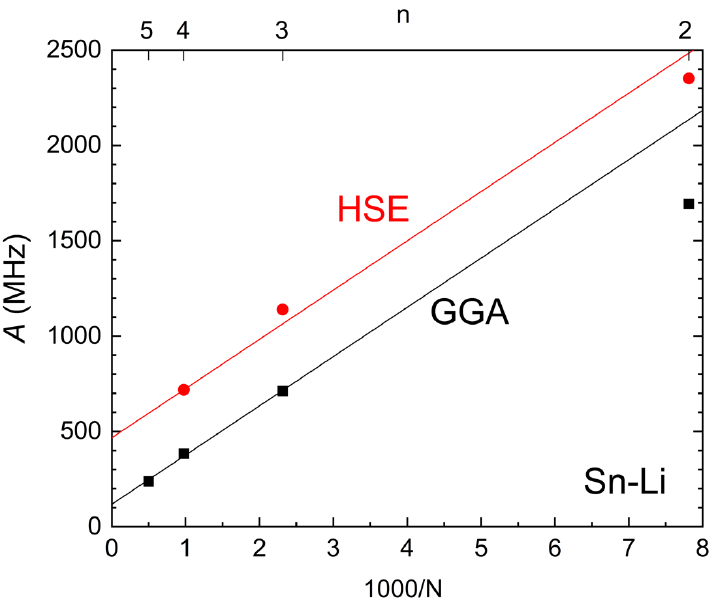}
    \caption{Isotropic hyperfine parameters for Sn-Li donors as a function of supercell size. GGA data points are in black squares and HSE data points are in red circles.The black line is a linear fit to GGA data for $N$ $\geq$ 432: $A$(GGA) = 117.99 + 258.26 $\times$ 1000/$N$. The red line is a rigid shift of the black line,  passing through the largest HSE data ($N$ $=$ 1024). The intercept of the red line yields $A= 466.7$~MHz.}
    \label{fig:hyperfine}
\end{figure}

We find that the hyperfine interactions are nearly identical for the different Sn–Li configurations. Therefore, we focus on a single configuration for the calculation of hyperfine parameters using large supercells.
The calculated isotropic hyperfine parameters are shown in Fig.~\ref{fig:hyperfine}.
A linear fit is first performed to the GGA results for supercells 
containing $N \geq 432$ atoms. Extrapolation to the dilute 
limit yields $A_{\mathrm{GGA}} = 118.0$~MHz. The fitted GGA line is 
then rigidly shifted to pass through the HSE data point obtained for 
the largest supercell. This procedure gives an extrapolated value 
$A_{\mathrm{HSE}} = 466.7$~MHz. To benchmark the approach, the same procedure is applied to calculate the hyperfine parameters of Ga and In shallow donors in ZnO. 
The resulting values are $A(^{69}\mathrm{Ga}) = 7.2$~MHz and 
$A(^{115}\mathrm{In}) = 103.0$~MHz, which are in agreement 
with the available experimental data~\cite{gonzalez1982magnetic}, summarized in Table~\ref{tab:hyperfine}.

Our calculated value for the Sn--Li complex, $A= 466.7$~MHz, confirms that this hyperfine parameter is exceptionally large for a shallow donor in ZnO, and is in acceptable agreement with the measured value of 392~MHz.

\begin{table}[t]
\caption{Calculated isotropic hyperfine parameters $A$ (in MHz) for $^{119}$Sn, 
$^{69}$Ga, and $^{115}$In, compared with available experimental values.~\cite{gonzalez1982magnetic} }
\label{tab:hyperfine}
\begin{ruledtabular}
\begin{tabular}{lccc}
 & Ga & In & Sn--Li \\
\hline
$A$ (Theory) & 7.2 & 103.0 & 466.7 \\
$A$ (Expt.)  & 18.8 & 102.5 & 388 (this work) \\
\end{tabular}
\end{ruledtabular}
\end{table}

\section{Conclusion and outlook}

We have demonstrated the formation and identification of the I$_{10}$ shallow donor in ZnO as a Sn–Li complex through a combination of ion implantation, annealing, optical spectroscopy, and first-principles calculations. Implantation experiments show that the I$_{10}$ emission emerges following Sn incorporation and is strongly enhanced by additional Li implantation. Two-laser coherent population trapping spectroscopy resolves a large electron–$^{119}$Sn hyperfine interaction of $A_{\mathrm{exp}} = 392 \pm 15$\,MHz and confirms an electron-spin-1/2, nuclear-spin-1/2 donor system. Density functional theory calculations further corroborate this identification, showing that a nearest-neighbor $\mathrm{Sn_{Zn}-Li_{Zn}}$ complex has favorable formation energetics, donor character with the electron localized on Sn, and an extrapolated hyperfine interaction in reasonable agreement with experiment. Together, these results establish the Sn–Li complex as the deepest shallow donor currently identified in ZnO with an exceptionally large hyperfine coupling, providing an attractive platform for spin-photon quantum technologies. 

The observed strong hyperfine interaction and large donor binding energy have several important implications for applications. The increased localization energy of the donor-bound exciton should enable improved thermal robustness of the optical transition relative to conventional group-III donors, while the large hyperfine interaction opens the possibility of fast electron–nuclear gates and potentially direct nuclear-spin–photon entanglement when the hyperfine splitting exceeds the optical linewidth. At the same time, the experiments reveal efficient optically driven nuclear spin polarization, indicating both an opportunity for rapid nuclear initialization and a potential challenge for maintaining nuclear spin coherence under optical excitation~\cite{Reiserer_2016}. 

Looking forward, several avenues can further advance this donor platform. Time-resolved studies of optical pumping dynamics and nuclear spin relaxation will clarify polarization mechanisms and enable optimized initialization protocols. Reduction of inhomogeneous broadening—through improved material growth, isotopic purification, or single-donor isolation—should allow direct optical resolution of the hyperfine structure. From a materials perspective, controlled co-implantation and annealing strategies may enable higher efficiency formation of Sn–Li complexes and integration into nanophotonic devices. More broadly, the methodology developed here for identifying and modeling complex shallow donors can be extended to other impurity pairs in ZnO and related wide-band gap semiconductors, potentially uncovering additional donor systems with enhanced hyperfine interactions and optical properties suitable for scalable quantum networks.

\begin{acknowledgments}
We acknowledge the very helpful discussion with Andrei Faraon, Yuan Ping, Juan Carlos Idrobo, and Simon Watkins. The authors thank Vasilieous Niaoris with help in initial CPT measurements. We thank Michael Titze and Edward Bielejec (Sandia National Labs) for initial Sn implantation trials. The work was primarly supported by the AFOSR CFIRE program
under grant FA9550-23-1-0418. Nuclear spin studies were supported by National Science Foundation under Grant No. 2212017. Theoretical calculations were supported by the Department of Energy, Office of Science, National Quantum Information Science Research Centers, Co-design Center for Quantum Advantage
(C2QA) under contract number DE-SC0012704. SIMS studies were supported by the Research Council of Norway through the project no. 325573. S.M. would like to acknowledge the startup fund from the University of South Carolina. 
The research used resources of the National Energy Research Scientific Computing Center, a DOE Office of Science User Facility supported by the Office of Science of the U.S. Department of Energy under Contract No. DE-AC02-05CH11231 using NERSC award BES-ERCAP0021021.

\end{acknowledgments}

\appendix

\renewcommand{\thefigure}{A\arabic{figure}}
\setcounter{figure}{0}

\section{Growth of MBE homoepitaxial ZnO}
\label{app:growth}
Homoepitaxial ZnO thin films are grown by molecular beam epitaxy detailed in\cite{Falson_2018}. Zn-polar single-crystal ZnO (0001) substrates [Tokyo Denpa] are chemically cleaned in an HCl solution with pH ranging from -0.5 to 1.5 \cite{Akasaka_2011}, followed by thermal degassing in a load-lock chamber at 200$^\circ$C for 1 h. The growth chamber is equipped with a turbo molecular pump, a liquid nitrogen shroud, and a custom-built CO$_2$ laser substrate heater \cite{Braun_2020, Kim_2025}, enabling a base pressure below 3$\times$10$^{-8}$ Pa at the growth temperatures. Elemental Zn [6N purity, Dowa electronics] is evaporated from an effusion cell operated at 280-300$^\circ$C. The Zn flux measured using a beam flux monitor is maintained at 5$\times$10$^{-4}$ Pa. Ozone gas is generated by evaporating pure liquid ozone [Meidensha] and delivered through a gas nozzle. The ozone flux is quantified using a capacitance manometer, which measured a local pressure of 100 mTorr behind the nozzle. The growth is carried out at a substrate temperature of 725$^\circ$C for 2 h. The background pressure during growth ranges from 1.0 to 3.0$\times$10$^{-5}$ Pa.

\section{\texorpdfstring{The excited Sn-Li donor bound exciton line: I$_{10}^*$}{The excited Sn-Li donor bound exciton line: I10*}}
\label{app:I10*}

\begin{figure}[h!]
    \centering
    \includegraphics[width=1\linewidth]{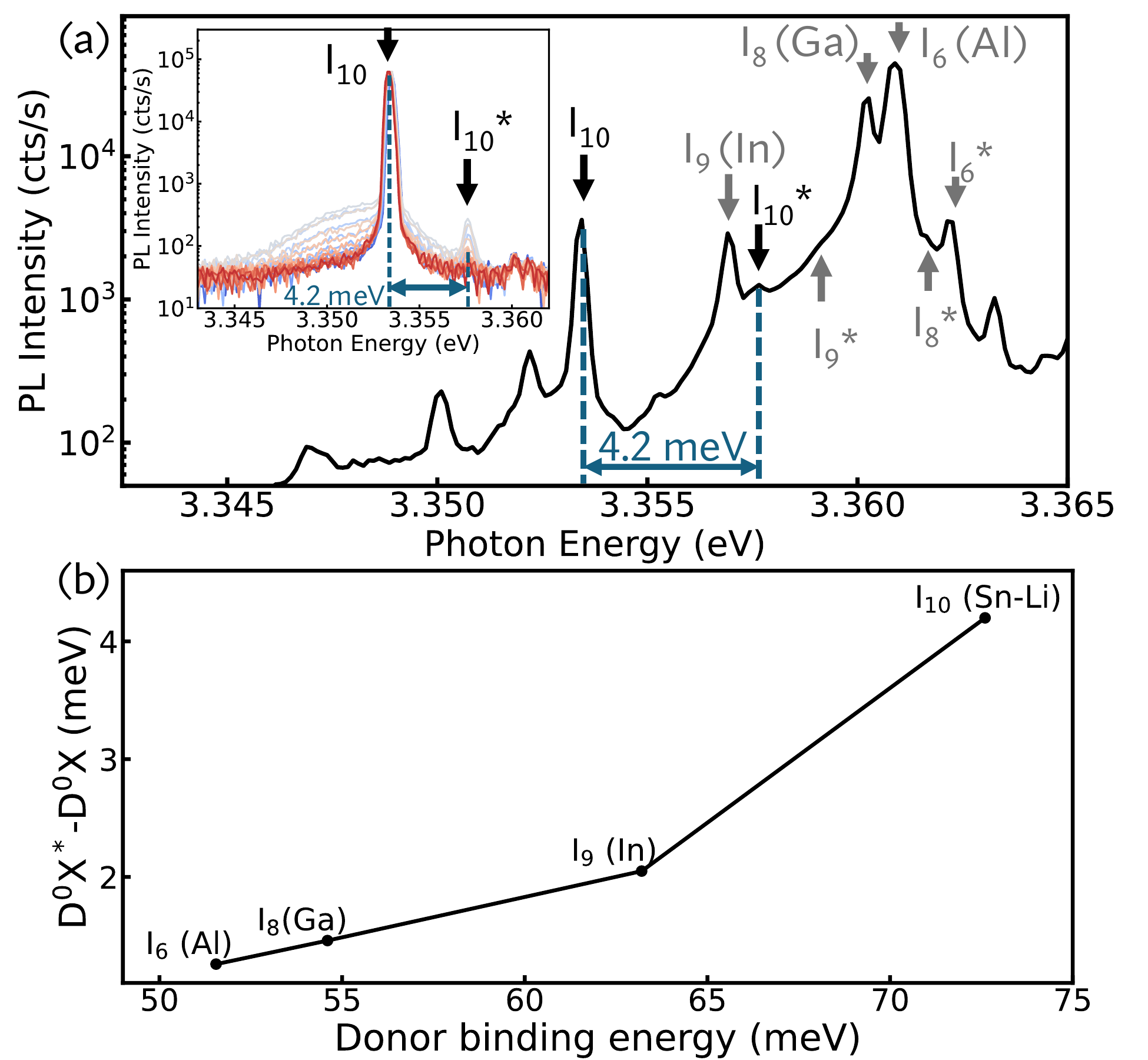}
    \caption{(a) PL of sample A at T=8\,K, excitation power 3.8$\mu$W at 360\,nm. The small peak labeled as I$_{10}^*$ is 4.2\,meV above the I$_{10}$ D$^0$X line. Inset: overlay of spectra sweeping across the resonance of I$_{10}$ with cross-polarized collection. (b) The energetic separation between D$^*$X of Al, Ga, In and Sn-Li donors with respect to donor binding energy in ZnO.}
    \label{fig:sn+}
\end{figure}

Figure \ref{fig:sn+}(a) shows a PL spectrum of shallow donors I$_{10}$(Sn-Li), I$_{9}$(In), I$_{8}$(Ga), and I$_{6}$(Al), together with their excited states D$^0$X$^*$ consistent with Ref.~\onlinecite{Niaouris_2024}. A small peak at 3.3582 eV (4.2\,meV above the main line of I$_{10}$) is attributed to the excited donor bound exciton line I$_{10}^*$. This attribution is further confirmed via resonant excitation on I$_{10}$ when a tunable laser sweeps across the main line of I$_{10}$ in the side excitation configuration\cite{Linpeng_2020} and cross-polarized PL is collected (inset of Fig.~\ref{fig:sn+}(a)). The intensity of the 4.2\,meV peak dramatically increases as the tunable laser becomes resonant with I$_{10}$ consistent with an excited state assignment. 

In comparison to Sn-Li, the corresponding excited states for Al, Ga, and In, are much closer to the main transitions, 1.26\,meV, 1.46\,meV, and 2.05\,meV, respectively, as illustrated in Figure \ref{fig:sn+}(b). The further the excited state is separated from the main transition, the less population relaxation to this state contributes to phonon-mediated temperature broadening of the main optical transition~\cite{Niaouris_2024}. For indium donors, there is negligible thermal contribution to the linewidth up to 5\,K. Analogously, the much larger energy difference observed for Sn-Li should enable higher temperature operation without thermal broadening.

\section{\texorpdfstring{I$_{10}$ Replicas}{I10 Replicas}}

\label{app:replicas}

\begin{figure}[h!]
    \centering
    \includegraphics[width=1\linewidth]{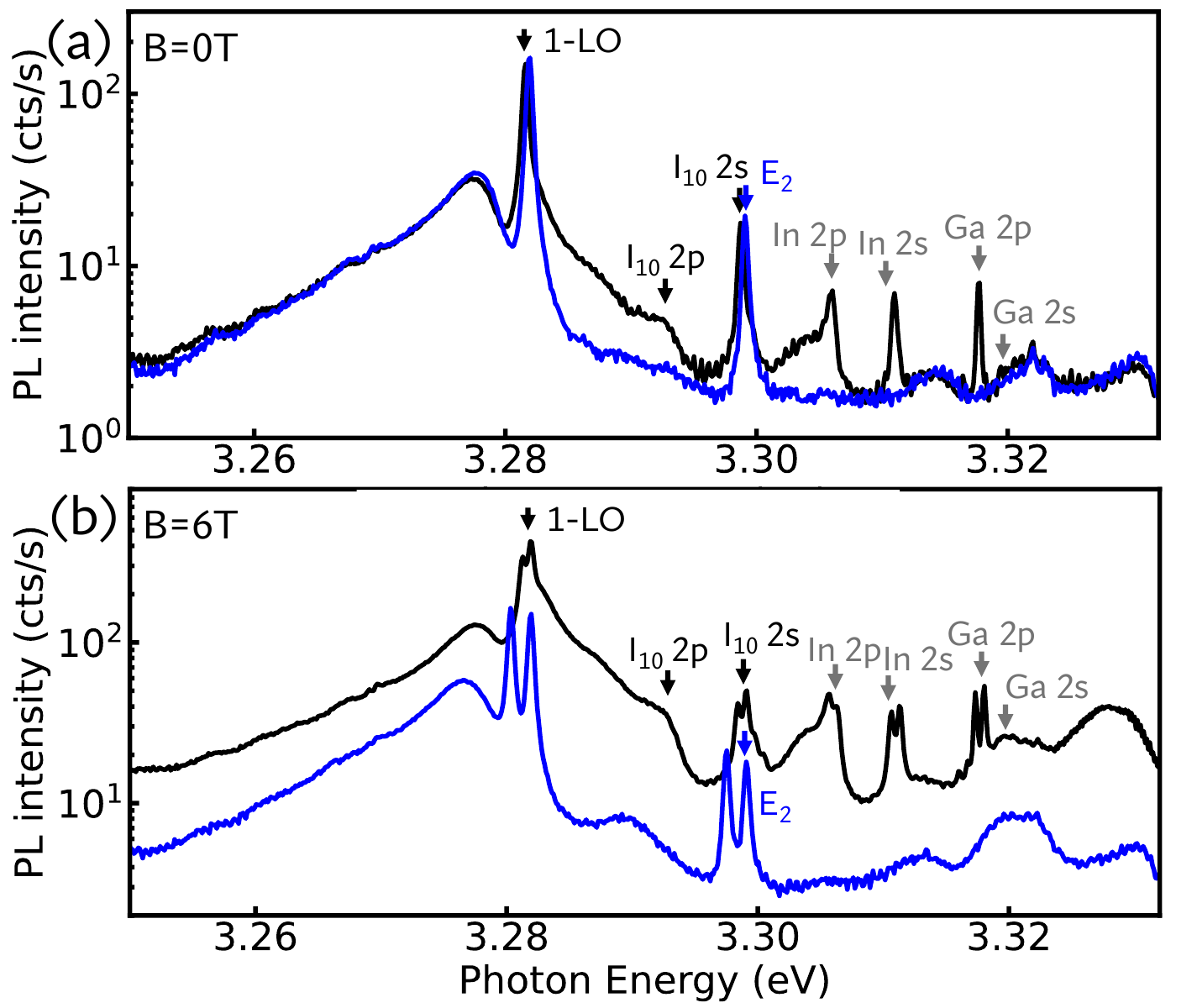}
    \caption{(a) Resonant excitation on/off I$_{10}$ transition at zero field. Resonant excitation energy is at 3.3532\,eV, off-resonant excitation energy is at 3.3535\,eV. T=8\,K, P$_\text{exc}$=5\,$\mu$W. (b) Resonant excitation with the pump beam on/off the low energy Zeeman transition of I$_{10}$ at 6\,T. E$_{\text{pump, on}}$=3.3529\,eV, E$_{\text{pump, off}}$=3.3519\,eV, E$_{\text{probe}}$=3.3535\,eV, P$_\text{pump}$=8\,$\mu$W, P$_\text{probe}$=5\,$\mu$W. T=8\,K.}
    \label{fig:TES}
\end{figure}

As shown in Figure \ref{fig:TES}(a), resonant excitation at zero field on I$_{10}$ enhances the two-electron satellite (TES) transitions from the donor bound exciton states to the excited 2s/2p donor state. 2s/2p of shallow donor. Raman scattering is also observed. The brightest feature is the 72\,meV 1-LO phonon Raman transition. Additionally, we observe strong scattering from the 54\,meV E$_2$(high) non-polar phonon mode \cite{Klingshirn_2007}. The Raman scattering is observed both on (black) and off (blue) resonance. At the instrument spectral resolution, the small detuning can be observed between the on- and off- resonance cases. The overlap of the E$_2$ line and the I$_{10}$ 2s line is coincidental. The mechanism for enhancement of the In and Ga TES transitions with resonant I$_{10}$ excitation is unexpected given the larger transition energy of the Al and In donor bound exciton. Additionally, these TES transitions are significantly sharper than any reported in the literature~\cite{Wang_2023}. This observed enhancement mechanism of this resonance is beyond the scope of this work. For this work, we observe that the PLE spectrum (e.g. as shown in Fig.~\ref{fig:CPT}) is identical when integrating over I$_{10}$ features or the Al/Ga features.  

At 6\,T, the resonant spectrum is taken at two-photon resonance condition illustrated in Fig.~\ref{fig:CPT}(a) to avoid optical pumping. The off-resonant spectrum is taken when the pump beam is detuned 0.9\,meV away from the two-photon resonance. This detuning is directly observed in the doublet of the Raman lines in the off-resonant spectrum. In Figure \ref{fig:TES}(b), the resonant spectrum (black) shows the Zeeman splitting of the I$_{10}$ 2s line as expected. The sharp features near the In and Ga TES transitions also show the characteristic electron-donor Zeeman splitting. 

\section{First-Principles Calculations}
\label{app:dft}
Density functional theory (DFT) calculations are performed using the projector augmented wave method (PAW)~\cite{Blochl_1994} implemented in the Vienna~\textit{ab-initio} simulation package (VASP)~\cite{Kresse_1993,Kresse_1996}. The PAW pseudopotentials for the host elements correspond to the valence-electron configuration 4$s^2$3$d^{10}$ for Zn and 2$s^2$2$p^4$ for O.  For the dopants, the electronic configurations are 2$s^1$ for Li, and 5$s^2$5$p^2$ for Sn. 
A plane-wave energy cutoff of 450 eV is employed throughout all calculations. 
To enable hyperfine interaction calculations for very large supercells required to obtain extrapolated results for shallow donors, we use the Generalized Gradient Approximation (GGA) of Perdew \textit{et al.}~\cite{Perdew_1996}.  Additionally, the hybrid functional of Heyd, Scuseria, and Ernzerhof (HSE) \cite{Heyd_2003}  is used to properly describe localization of the wave function, with a mixing parameter $\alpha=0.36$. This produces a band gap of 3.31 eV, comparing well with the experimental value of 3.44 eV~\cite{jagadish2011zinc}. 
Note that we compare the hyperfine parameters obtained from $\alpha=0.36$ and $\alpha=0.4$, and the difference is negligible. 
For the primitive cells, full structural optimizations are performed using an 8$\times$8$\times$8 $K$-point mesh. 
For both bulk and supercells, atomic positions are optimized until the Hellmann-Feynman forces are lower than 5 meV/\AA.  

The primitive four-atom hexagonal cell of ZnO is not suitable for the defect calculation, as it does not provide comparable impurity separations along the three orthogonal directions. Therefore, we apply the following lattice transformation to connect the new lattice parameters $a^\prime$, $b^\prime$, $c^\prime$ to lattice vectors $a$, $b$, $c$ of a hexagonal 4-atom primitive cell.  
\begin{equation}
\begin{pmatrix}
a^\prime \\
b^\prime \\
c^\prime
\end{pmatrix}
= \begin{pmatrix}
2 &  0 & 0 \\
1 &  2 & 0 \\
0 &  0 & 1
\end{pmatrix}
\begin{pmatrix}
a \\
b \\
c 
\end{pmatrix}
.
\end{equation} 
This transformation yields a 16-atom orthorhombic cell with comparable lattice parameters $a^\prime = 6.481\,\text{\AA}$, $b^\prime= 5.613\,\text{\AA}$, and $c^\prime = 5.224\,\text{\AA}$. 
The $n \times n \times n$ supercells constructed from this orthorhombic cell provide similar separation between the impurity and its periodic images along the three Cartesian directions. 

The smallest supercell ($n = 2$) contains $N = n \times n \times n \times 16 = 128$ atoms, which serves as the simulation cell for the defect formation energy calculations. 
The formation energy of a single cation substitutional defect $X$ in charge state $q$ is given by~\cite{Freysoldt_2014}:
\begin{equation}
\begin{split}
E^f\left[X^q\right]=E_{\mathrm{tot}}\left[X^q\right]-E_{\mathrm{tot}}[\text { bulk }]-\mu_X \\
+\mu_\text{Zn}+q E_{\mathrm{F}}+\Delta^q.
\end{split}
\end{equation}
Here, $E_{\mathrm{tot}}[\text { bulk }]$ and $E_{\mathrm{tot}}\left[X^q\right]$ represent the total energies of a bulk supercell and a supercell with the defect, respectively.  A special $k$ point (0.25, 0.25, 0.25) was used for Brillouin-zone (BZ) integration to calculate these energies. 
The chemical potential for element $i$ is denoted as $\mu_i$. 
The Fermi level, or the electron chemical potential, is denoted by $E_{\mathrm{F}}$ and is referenced to the valence-band maximum (VBM). The term $\Delta^q$ is a finite-size correction for the electrostatic interaction between a charged defect and its images~\cite{Freysoldt_2009,Freysoldt_2011}.  
For Li incorporation, we use Li$_2$O$_2$ and Li metal as the limiting phases under O-rich and Zn-rich conditions, respectively.  
For Sn incorporation, we use SnO$_2$ and Sn as the limiting phases under O-rich and Zn-rich conditions, respectively. 

The charge-state transition level between charge states $q$ and $q'$, denoted as ($q/q'$), is calculated based on the formation energies:
\begin{equation}\label{eq_levels}
    (q/q') =\frac{E^f(\text{X}^q_\text{Zn}; E_\text{F}=0)-E^f(\text{X}^{q'}_\text{Zn}; E_\text{F}=0)}{(q'-q)} \, .
\end{equation}

To calculate the hyperfine parameter for the shallow donor state, we adopt the approach developed in Swift \emph{et al.}~\cite{swift2020first,karbasizadeh2024transition}. In this method, a series of $n \times n \times n$ supercells is constructed as described above and the hyperfine parameter is extrapolated to the dilute limit. For the GGA calculations, supercells ranging from $n = 2$ (128 atoms) to $n = 5$ (2000 atoms) are considered.
For the more computationally demanding HSE functional, $n = 4$ (1024 atoms) represents our practical upper limit. For these supercell calculations of the hyperfine parameters, BZ integrations are carried out using the $\Gamma$ point as the special point; we have tested that this is sufficient for the large supercells employed here.  

\bibliography{main.bib}

\end{document}